\documentclass[12pt]{article}
\usepackage[english]{babel}
\usepackage[latin1]{inputenc}
\usepackage{amsmath}
\usepackage{graphicx,psfrag,epsf}       
\usepackage{amsthm}                     
\usepackage{amsfonts}                   
\usepackage{bbm}                        
\usepackage{natbib}                     
\usepackage{mathtools}                  
\mathtoolsset{showonlyrefs}             
\usepackage[title, titletoc]{appendix}  
\usepackage{booktabs}
\usepackage{dcolumn}
\usepackage{tikz}
\usepackage{pgfplots}
\usepackage{pgfplotstable}
\usepackage{filecontents}
\usepackage[colorlinks=true,linkcolor=blue,citecolor=blue]{hyperref}
\usepackage{algorithm}
\usepackage{color}
\usepackage{enumerate}                  
\usepackage{setspace}                  
\usepackage[bf]{caption}                
\usepackage[normalem]{ulem}             
\usepackage{longtable}                  
\usepackage{booktabs}

\allowdisplaybreaks                     

\usepgfplotslibrary{groupplots} 

\newcolumntype{.}{D{.}{.}{1.6}}
\newcolumntype{:}{D{.}{.}{1.2}}
\newcolumntype{;}{D{.}{.}{5.2}}

\def\xb{{\bf x}}

\def\Pr{{\rm Pr}}
\def\E{{\rm E}}

\def\V{{\rm V}}

\def\betab{\boldsymbol{\beta}} 

\def\1{\mathbbm 1}

\def\R{\mathbb R}

\floatname{algorithm}{Procedure}

\title{Nonparametric imputation method for nonresponse in surveys}%
\author{Caren Hasler\footnote{Affiliation while the research was conducted: Institute of Statistics, University of Neuch\^atel, Av. de Bellevaux 51, 2000 Neuch\^atel, Switzerland. Current affiliation: Department of Computer and Mathematical Sciences, University of Toronto Scarborough, 1265 Military Trail, Toronto, Ontario, M1C 1A4, Canada.}
~~and
Radu V. Craiu\footnote{Department of Statistical Sciences, University of Toronto, 100 St. Georges Street, Toronto, Ontario, M5S 3G3, Canada}}

\begin{document}

\maketitle

\begin{abstract}
    Many imputation methods are based  on statistical models that assume that the variable of interest is a noisy observation of a function of the auxiliary variables or covariates. Misspecification of this model may lead to severe errors in estimates and to misleading conclusions. A new imputation method for item nonresponse in surveys is proposed based on a nonparametric estimation of the functional dependence between the variable of interest and the auxiliary variables.
    We consider the use of smoothing spline estimation within an additive model framework to flexibly build an imputation model in the case of multiple auxiliary variables. The performance of our method is assessed via numerical experiments involving simulated and real data.
\end{abstract}


\section{Introduction}

Nonresponse in surveys is a commonly encountered problem that, when ignored, can affect the performance of the statistical estimators for the quantities of interest.  Two general adjustment techniques that have been developed to alleviate the effects  of nonresponse are \emph{reweighting} and \emph{imputation}. Reweighting procedures consist of increasing the initial weights of respondents in order to compensate for nonrespondents and are commonly used to treat unit nonresponse. Imputation procedures consist of filling in the missing values in the data with \emph{imputed values} and are commonly used to treat item nonresponse. When dealing with nonresponse, both reweighting and imputation may rely on a statistical model. Imputation for the variable of interest can be more efficient if it is based on information contained in a number of auxiliary variables, specifically, through a model that estimates a functional link between the latter and the variable of interest. However, the validity of the model will have a direct effect on the accuracy of the estimated quantities. It is therefore crucial to be able to build flexible models that can capture a large spectrum of patterns and make only weak assumptions about the true underlying mechanism generating the data. Given these constraints, it is not surprising that nonparametric models have been used to handle nonresponse in surveys.

\cite{gio:87} focused on unit nonresponse and proposed two nonparametric reweighting procedures based on kernel density estimators to estimate response probabilities. Later, \cite{niy:94,niy:97} used the nonparametric estimation of~\cite{gio:87} to handle nonresponse when unit nonresponse and item nonresponse occur together. Finally, \cite{das:ops:06} and \cite{das:ops:09} applied, respectively, kernel regression and local polynomial regression to estimate the response probabilities and derived asymptotic properties of the propensity score adjusted estimator for these approaches. These techniques are suitable when the number of auxiliary variables is relatively low.

We propose here an imputation method for item nonresponse in surveys when the variable of interest is a noisy observation of a function of many auxiliary variables. We consider smoothing spline models within an additive regression framework which allows us to handle a large number of auxiliary variables. This improvement significantly expands the range of  nonparametric methods for handling nonresponse. Moreover, the model considered is adaptable to a wide variety of functional patterns thus providing protection against model misspecification. Results of a simulation study confirm the performance of our method and highlight its capacity to adapt to many different situations.

The paper is organized as follows: Section~\ref{section:framework} establishes the framework and introduces notation; Section~\ref{section:motivation} provides a motivation for the new imputation method; two nonparametric tools used in the new imputation method are reviewed in Section~\ref{section:nonparametric}; Section~\ref{section:method} presents the new method as well as bootstrap procedures to estimate the variance of the total. The performance of the new method is compared to that of other imputation methods through a simulation study presented in Section~\ref{section:simulation}. We close with concluding remarks and a discussion of future work.

\section{Framework}\label{section:framework}

Consider a finite population $U = \left\{ 1, 2, \ldots, N \right\}$ of possibly unknown size $N$. Suppose that the parameter of interest is the population total
\begin{align}\label{equation:Y}
    Y = \sum_{i \in U} y_i,
\end{align}
for some unknown variable of interest $y$. A sample $S$ of size $n$ is selected from $U$ according to a probabilistic sampling design $p(\cdot)$ with the aim of observing $y_i$ for $i \in S$. Consider
\begin{align}
    \pi_i = \Pr(i \in S) = \sum_{s \subset U;s \ni i}p\left( s \right),
\end{align}
the first-order inclusion probability of unit $i$ and suppose that $\pi_i > 0$ for all $ i \in U$. Let $d_i= 1/\pi_i$ represent the design weight of unit $i \in U$. In this paper we consider two widely used  sampling designs,  simple random sampling without replacement (SRSWOR) and stratified sampling (SS). Under SRSWOR, each sample of (fixed) size $n$ has the same probability of being selected and $\pi_i = n/N$ for all $i \in U$. Under SS, the population $U$ is partitioned into $H$ strata $U_1, \ldots, U_H$ of respective sizes $N_1,\ldots, N_H$ and SRSWOR is applied independently in each stratum $h$. A sample $S_h$ of size $n_h$ is hence selected in each stratum $U_h$, $h = 1, \ldots, H$ and $\pi_i = n_h/N_h$ for all $i \in U_h$.

Once a sample $S$ is selected, each unit $i \in S$ is classified as either respondent or nonrespondent, depending on whether $y_i$ is observed or missing. Consider the response indicator vector $\left( r_i | i \in S\right)^\top$ where $r_i$ takes value 1 if $y_i$ is observed and 0 if it is missing. This results in the set of respondents $S_r = \left\{ i \in S | r_i = 1 \right\}$ and in the set of nonrespondents $S_m = \left\{ i \in S | r_i = 0 \right\}$.

Under complete response, the Horvitz-Thompson estimator
\begin{align}\label{equation:ht}
    \widehat{Y} = \sum_{i \in S} \frac{1}{\pi_i} y_i,
\end{align}
is a design unbiased estimator for $Y$, i.e. $\E_p(\widehat{Y}) = Y$. 
In the case of a survey with nonresponse, however, the estimator~\eqref{equation:ht} cannot be computed since some of the $y_i$'s, $i \in S$ are missing. One remedy is to impute each missing value $y_i$, $i \in S_m$ with an imputed value $y_i^*$. The population total $Y$ can then be estimated through the {\it imputed estimator}
\begin{align}\label{equation:imputed:estimator}
    \widehat{Y}_I = \sum_{i \in S} \frac{1}{\pi_i} \left[ y_i r_i + y_i^* \left(1 - r_i\right) \right]
    = \sum_{i \in S_r} \frac{1}{\pi_i} y_i + \sum_{i \in S_m} \frac{1}{\pi_i} y_i^* = \sum_{i \in S} \frac{1}{\pi_i} \widetilde{y}_i,
\end{align}
where
\begin{align}
    \widetilde{y}_i = \left\{
                        \begin{array}{ll}
                          y_i & \hbox{if $i \in S_r$;} \\
                          y_i^* & \hbox{if $i \in S_m$.}
                        \end{array}
                      \right.
\end{align}
If the imputation process exactly reconstructs the missing values, that is if $y_i^* = y_i$ for $i \in S_m$, then $\widehat{Y}_I$ is a design unbiased estimator for the population total $Y$. Hence, an imputation method that reconstructs the missing data well can provide protection against nonresponse bias. Design weights can optionally be taken into account when constructing the imputed values, the resulting method being referred to as \emph{survey weighted imputation}.

Consider a vector $\xb_i = \left( x_{i1}, x_{i2}, \ldots, x_{iq} \right)^\top$ of values taken by $q$ auxiliary variables $x_1$, $x_2$, $\ldots$, $x_q$ and known for all $i \in U$ or at least for all $i \in S$. Auxiliary information can be used at different stages of the survey, namely in establishing the sampling design, for estimation, and handling of nonresponse. Reliable auxiliary information can explain the variation in the variable of interest and/or in the response probabilities and helps reduce error due to sampling and nonresponse.

\section{Motivation}\label{section:motivation}

We consider a variable of interest, $y$, that is measured along with $q$ auxiliary variables, $x_{1},\ldots, x_{q}$. In situations in which the variable of interest is not recorded for some sampled units, one may rely on the auxiliary variables to impute the missing values if there is a way to connect these variables via an \emph{imputation model}~\citep{sar:92}. For instance, consider a general model of the type
\begin{align}\label{imp:mod:0}
    y_i = f(x_{i1}, x_{i2},\ldots,x_{iq}) + \varepsilon_i,
\end{align}
where $f$ is a function from $\R^q$ to $\R$, and $\varepsilon_i$ are zero-mean independent errors with variance $\sigma^2$. A deterministic imputation method estimates  first the function $f$ based on those individuals/items $i \in S_r$ for which $(y_i, \xb_i) = (y_i, x_{i1},\ldots,x_{iq})$ are fully observed, and then imputes values for $i \in S_m$ using the estimated function and the observed $\xb_i$. The challenging issue of estimating $f$ naturally arises because the choice of the imputation model crucially impacts the accuracy of the imputed values. A misspecified model may result in highly biased estimates for the parameters of interest.

Without prior knowledge on the form of $f$ in \eqref{imp:mod:0}, it is natural to use a nonparametric regression model since the resulting estimate $\hat f$ is known to adapt to the shape of $f$ based on the information provided by the data. When handling survey data, however, several  auxiliary variables are often available and one needs to include most of them in the model. Unfortunately, a few nonparametric smoothers such as kernel-based ones  tend to break down in high dimension, unless the sample size is very large. This phenomenon is known as the \emph{curse of dimensionality} \citep{bel:61,sto:85} and can be alleviated if an additive model \citep[AM,][]{has:tib:86} is used. Such a model is additive in the predictor variables and takes the form
\begin{align}\label{imp:mod}
     y_i = a_0 + \sum_{j = 1}^q a_j(x_{ij}) + \varepsilon_i,
\end{align}
where $(y_i, \xb_i) = (y_i, x_{i1},\ldots,x_{iq})$, $i=1,\ldots,N$, are observations, $a_0$ is a constant, $a_j$, $j = 1,\ldots,q$, are univariate smooth functions, and $\varepsilon_i$ are zero-mean independent errors with common variance $\sigma^2$. The functions $a_j$, $j = 1,\ldots,q$, are each individually estimated by univariate smoothers so the curse of dimensionality is avoided because the original problem of nonparametric estimation in $\R^{q}$ has been replaced by $q$ estimation problems in $\R$.
Without loss of generality, henceforth we suppose that the $\xb_i$, $i = 1,\ldots,N$, lie in the interval $[0,1]^{q}$.

We propose an imputation method for nonresponse in survey based on AM. The new method is based on imputation model \eqref{imp:mod}. The nonparametric tools used to estimate the regression function are presented in Section~\ref{section:nonparametric} and the new method is presented in Section~\ref{section:method}.

\section{Nonparametric tools}\label{section:nonparametric}


This section introduces two nonparametric tools used in the new imputation method, smoothing spline regression and additive models. The main idea of smoothing spline regression is to fit a data set with a curve that maximizes a measure of goodness-of-fit while achieving a fixed degree of smoothness. There is an extensive literature devoted to spline regression and we refer the reader to \cite{gre:sil:94}, \cite{eub:99}, and \cite{wan:11}.
Smoothing spline regression (SSR) assumes model \eqref{imp:mod} with a unique predictor variable, that is
\begin{align}
    y_i = a(x_i) + \varepsilon_i, \;\; 1\le i \le N
\end{align}
where $\varepsilon_i$ are zero-mean independent errors with common variance $\sigma^2$, and $a$ is a smooth function in the sense that $a \in W_2^m[0,1]$ where $W_2^m[0,1]$ is the Sobolev space
\begin{align}
    W_2^m[0,1] = \left\{ g: g,g',\ldots,g^{(m-1)} \mbox{ are absolutely continuous, } \int_{0}^1 g^{(m)}(t)^2 < +\infty  \right\} .
\end{align}
We consider a  basis of functions $b_k$, $k \in 1, \ldots K$, called \emph{spline basis functions}, for $W_2^m[0,1]$.
The SSR yields the best approximation of function $a$ in $W_2^m[0,1]$ while controlling the degree of smoothness. The resulting \emph{smoothing spline estimator} $\widehat{a}$  is the minimizer of the following penalized least square (PLS) criterion
\begin{align}\label{equation:pls}
    \frac{1}{N} \sum_{i = 1}^N \left( y_i - g(x_i) \right)^2 + \lambda \int_{0}^1 g^{(m)}(t)^2 dt,
\end{align}
over all functions in $W_2^m[0,1]$. The parameter $\lambda$ is the \emph{smoothing parameter} and its size decides the balance between goodness-of-fit, as measured by the mean squared residual, and smoothness, as measured by the integral. There exist different basis of functions, each of which can produce a different smoothing spline estimator. In what follows, we will consider the thin plate spline basis~\citep[see][]{woo:03} and the smoothing parameter $\lambda$ will be selected by generalized cross validation.

With survey data, it is often desirable to consider design weights when estimating parameters of interest. Indeed, a design weight $d_i = 1/\pi_i$ can be interpreted as the number of population units that sampled unit $i$ represents. Hence, when units are selected with unequal inclusion probabilities  it might be unreasonable to assume that each sampled unit has the same influence on the parameters of interest.
A weighted version of the smoothing spline estimator was proposed by~\cite{zha:chr:zhe:13} who suggested adding design weights in the general PLS criterion in equation~\eqref{equation:pls}. Hence, they consider the smoothing spline estimator adapted for survey data which is the minimizer over $g$ of
\begin{align}\label{equation:pls:pi}
    \frac{1}{\widehat{N}} \sum_{i \in S} d_i \left( y_i - g(x_i) \right)^2 + \lambda \int_{0}^1 g^{(m)}(t)^2 dt,
\end{align}
where $\widehat{N} = \sum_{i \in S} d_i$ is the estimated population size. Note that \cite{zha:chr:zhe:13} restrict themselves to the case $m = 2$.

A flexible way to combine the contributions of each auxiliary variable to the variable of interest is provided by the additive model paradigm.  A class of generalized additive models was proposed by \cite{has:tib:86} and was discussed in depth in the book~\cite{has:tib:90}. We focus here on the additive regression model (AM), which assumes
\begin{align}\label{model:gam}
    y_i = a_0 + \sum_{j = 1}^q a_j(x_{ij}) + \varepsilon_i,
\end{align}
where $a_0$ is a constant, $a_j$, $j = 1, \ldots, q$, are smooth functions, and $\varepsilon_i$ are zero-mean independent errors with common variance $\sigma^2$. SSR is used to estimate each function $a_j$, $j = 1, \ldots, q$. A backfitting  algorithm \citep{has:tib:86} or a direct fitting approach \citep{woo:08} can be considered.

When appropriate, an additive model allows us to handle multiple predictor variables in a reasonable computation time and avoids the curse of dimensionality problem as it breaks a high-dimensional nonparametric estimation problem into a number of one-dimensional ones.

\section{The method}\label{section:method}

In this section, we propose a nonparametric model-based imputation method for nonresponse in surveys and discuss bootstrap procedures to estimate the resulting  variance of the  total estimator for the population $U$.

\subsection{Estimation and imputation}

Assume that the sample $S$ contains respondents $S_r$ for which the values of the variable of interest $\{y_{i}:\; i\in S_{r}\}$ are observed and nonrespondents for which these values $\{y_i:\; i \in S_m\}$ are missing. For each unit $i \in S$ we have available auxiliary variables values $\xb_{i}=\{x_{i1},\ldots,x_{iq}\}$. We consider the following additive imputation model
\begin{align}\label{model:imputation:am}
    y_i = a_0 + \sum_{j = 1}^q a_j(x_{ij}) + \varepsilon_i,
\end{align}
where  $a_0$ is a constant, $a_j$, $j = 1,\ldots,q$, are univariate functions in the functional space defined in Section~\ref{section:nonparametric}, and $\varepsilon_i$ are zero-mean independent errors with common variance $\sigma^2$. Smoothing spline estimates $\widehat{a}_j$, $j = 1, \ldots,q$, of functions $a_j$, $j = 1, \ldots,q$, and an estimate $\widehat{a}_0$ of $a_0$ are obtained using the complete data $(y_i, \xb_i)$, $i \in S_r$. Two different smoothing splines estimators can be obtained based on expression \eqref{equation:pls} (unweighted imputation) or expression~\eqref{equation:pls:pi} (survey weighted imputation), respectively. Finally, missing values $y_i$, $i \in S_m$, are imputed with predictions based on imputation model~\eqref{model:imputation:am} as follows
\begin{align}\label{imputed:values}
    y_i^* = \widehat{a}_0 + \sum_{j = 1}^q \widehat{a}_j(x_{ij}).
\end{align}

\subsection{Variance estimation for the imputed total}\label{section:variance}

A valid method for estimating the variance of the  estimator of the population total must account for the extra variability due to imputing the missing values. In turn, this variability is due to the variance of predicted values $y_{i}^{*}$ produced via the additive model. Since an analytical expression for the asymptotic error of AM predictive value is not available, we pursue a bootstrap-based approach. Bootstrap procedures to estimate the variance of parameters of interest are available for different imputation methods and sampling designs. In this Section, we follow \cite{sha:sit:96} to devise bootstrap procedures to estimate the variance of the total under AM imputation for simple random sampling without replacement (SRSWOR) and stratified sampling (SS). The bootstrap proposed in \cite{sha:sit:96} is asymptotically valid irrespective of the sampling design, or the imputation method.

We follow \cite{sha:sit:96} and apply the without-replacement bootstrap (BWO) proposed by~\cite{gro:80} to estimate the variance of the total under AM imputation for SRSWOR. Procedure~\ref{procedure:1} presents the applied procedure which proceeds as follows. Given a sample of size $n$ from a population of size $N$,  we set $k = N/ n$ and assume $k$ is an integer (otherwise we round it off). In step 1 we  construct a pseudopopulation of size $N$ by replicating the sample $k$ times.  In step 2, a simple random sample of size $n$ is selected from the pseudopopulation. Because the pseudopopulation consists of sampled units, the bootstrap sample is very likely to contain both units with missing $y_i$ and units with observed $y_i$. In step 3, AM imputation is applied to the bootstrap sample. Steps 2 and 3 are repeated to obtain $B$ analogs of the imputed total estimator. In step 5, the bootstrap variance of the imputed total is obtained using the standard bootstrap formulae.

\begin{algorithm}[!htb]
\caption{Variance of the imputed total estimator under SRSWOR.\label{procedure:1}}
\begin{enumerate}[{Step} 1:]
    \item   Suppose $N = kn$ for an integer $k$.\\
            Construct a pseudopopulation by replicating the sample $k$ times.
    \item   Draw a SRSWOR of size $n$ from the pseudopopulation of step 1.
    \item   Apply AM imputation to impute the missing $y_i$'s of the sample selected in step 2.
    \item   Repeat steps 2 and 3 a large number of times $B$ to obtain $\widehat{Y}_I^{(1)},\ldots,\widehat{Y}_I^{(B)}$ where
            $\widehat{Y}_I^{(b)}$ is the analog of $\widehat{Y}_I$ for the $b$-th bootstrap sample.
     \item Obtain the bootstrap variance of $\widehat{Y}_I$ by
            \begin{align}
                \V_{boot}(\widehat{Y}_I) = \frac{1}{B} \sum_{b = 1}^B \left( \widehat{Y}_I^{(b)} - \widehat{Y}_I^{(.)} \right)^2,
            \end{align}
            where $\widehat{Y}_I^{(.)}$ is the mean bootstrap analog of $\widehat{Y}_I$
            \begin{align}
                \widehat{Y}_I^{(.)} = \frac{1}{B}\sum_{b = 1}^B \widehat{Y}_I^{(b)}.
            \end{align}
\end{enumerate}
\end{algorithm}

For SS, we also follow \cite{sha:sit:96} and apply the mirror-match bootstrap (MMB) proposed by~\cite{sit:92b} to estimate the variance of the total under AM imputation. Procedure~\ref{procedure:2} presents the applied procedure. In steps 1 and 2, the procedure mimics the stratified sampling by selecting several times SRSWOR of size $n_h'$ in stratum $h$. If $n_h'$ is such that $n_h' = f_h n_h$, then the size of the bootstrap sample $S_h^*$ is the same as that of $S_h$, i.e. $n_h^* = n_h$. This procedure is repeated independently in each stratum $h$ times to obtain a bootstrap sample ${\bf S}^*$. Because the bootstrap sample consists of sampled units, it is very likely to contain both units with missing $y_i$ and units with observed $y_i$. Hence, in step 4, AM imputation is applied to the bootstrap sample ${\bf S}^*$ and the bootstrap analog $\widehat{Y}_I^{(b)}$ of the imputed total estimator $\widehat{Y}_I$ is obtained. Depending on the choice of $n_h'$ and on whether randomization is applied to round $n_h'$ and/or $k_h$, the bootstrap procedure might mimic a stratified sampling in a population whose size differs from $N$. Fraction $N/n^*$ appears in the computation of the bootstrap analog of the imputed total estimator $\widehat{Y}_I$ to take this into account. Steps 1 to 4 are repeated to obtain $B$ analogs of the imputed total estimator. In step 6, the bootstrap variance of the imputed total is obtained using the standard bootstrap formulae.

\begin{algorithm}[!htb]
\caption{Variance of the imputed total estimator under SS.\label{procedure:2}}
\begin{enumerate}[{Step} 1:]
    \item  Choose $1 \leq n_h' < n_h$ and select a SRSWOR of size $n_h'$ without replacement from $S_h$.\\
             If $n_h'$ is not integer, apply a randomization~\citep[see][]{sit:92b}.
    \item   Repeat step 1 $k_h = n_h  (1 - f_h^*) / ( n_h' (1 - f_h) )$ times independently to obtain a sample $S_h^* = \left\{ hi: i = 1,\ldots, n_h^*\right\}$ of size $n_h^* = n_h' k_h$,
            where $f_h = n_h / N_h$ and $f_h^* = n_h' / n_h$.\\
            If $k_h$ is not integer, apply a randomization~\citep[see][]{sit:92b}
    \item   Repeat steps 1 and 2 independently for each stratum $h$ to obtain a bootstrap sample ${\bf S}^* = \left\{ S_1^*,\ldots,S_H^* \right\} = \left\{ hi: h = 1, \ldots, H; i = 1, \ldots, n_h^* \right\}$ of size $n^* = \sum_{h = 1}^H n_h^*$.
    \item   Apply AM imputation to impute the bootstrap sample ${\bf S}^*$ and obtain the bootstrap analog of the imputed total estimator $\widehat{Y}_I$ by
            \begin{align}
                \widehat{Y}_I^{(b)} = \frac{N}{n^*} \sum_{hi \in {\bf S}^*} \frac{\widetilde{y}_{hi}^{(*)}}{f_h^*}
                    = \frac{N}{n^*} \sum_{h = 1}^H \frac{n_h}{n_h'} \sum_{hi \in S_h^*} \widetilde{y}_{hi}^{(*)} ,
            \end{align}
            where $\widetilde{y}_{hi}^{(*)}$ is the value of the variable of interest of unit $hi$ if this one is observed and the imputed value otherwise.
    \item   Repeat steps 1 to 4 a large number of times $B$ to obtain $\widehat{Y}_I^{(1)},\ldots,\widehat{Y}_I^{(B)}$ where $\widehat{Y}_I^{(b)}$ is the analog of $\widehat{Y}_I$ for the $b$-th bootstrap sample.
    \item Obtain the bootstrap variance of $\widehat{Y}_I$ by
            \begin{align}
                \V_{boot}(\widehat{Y}_I) = \frac{1}{B} \sum_{b = 1}^B \left( \widehat{Y}_I^{(b)} - \widehat{Y}_I^{(.)} \right)^2,
            \end{align}
            where $\widehat{Y}_I^{(.)}$ is the mean bootstrap analog of $\widehat{Y}_I$
            \begin{align}
                \widehat{Y}_I^{(.)} = \frac{1}{B}\sum_{b = 1}^B \widehat{Y}_I^{(b)}.
            \end{align}
\end{enumerate}
\end{algorithm}

The computational time involved in the bootstrap evaluation of variance can be shortened if multiple processors are available. The embarrassing parallel
structure of the procedure implies that the sample-specific calculation can be performed on a separate processor and the merging of simulated values is needed  only in Step 5 (for Procedure 1) and in Step 6 (for Procedure 2).

\section{Simulations}\label{section:simulation}

A numerical study was conducted to test the performance of the proposed imputation method. Simulated data and real data were considered. In Sections~\ref{section:setting:1} and~\ref{section:setting:2}, the simulation settings for the simulated data and for the real data are respectively presented. Measures used to compare the new imputation method with existing imputation methods and to test the accuracy of the bootstrap procedures for the variance estimation are described in Section~\ref{section:comparison}. Finally, the results of the simulations in each setting are displayed and commented in Sections~\ref{section:results:setting:1} and \ref{section:results:setting:2} respectively.

\subsection{Setting 1: simulated data}\label{section:setting:1}

Populations of size $N = 10000$ were considered. Four auxiliary variables $x_1$, $x_2$, $x_3$, and $x_4$ were generated. The values $x_{i1}$, $x_{i2}$, and $x_{i3}$, $i = 1, \ldots, N,$ are independent draws from a $\mbox{Uniform}[0,1]$ random variable and  $x_{i4}$, $i = 1, \ldots, N,$ are independent draws of a gamma density with shape and scale parameters, respectively, 3 and 1/6 that were mapped into  the $[0,1]$ interval via the transformation $x_{i4} \rightarrow \left(x_{i4} - \min(x_4)\right)/\left(\max(x_4) - \min(x4)\right)$.

Five populations were then generated as follows:
\begin{align}
    y_i^{(1)} &= 1 + 5 x_{i1} + x_{i2} + x_{i3} + x_{i4} + \varepsilon_i,\\
    y_i^{(2)} &= 2 + \cos(\pi x_{i1} + \pi) + \sin (4\pi x_{i2}) + \exp(-(x_{i3} - 0.5)^2) + (x_{i4} - 0.5)^2 + \varepsilon_i,\\
    y_i^{(3)} &= 1 + \cos(2\pi x_{i1}) + x_{i1}x_{i2} + x_{i3}^2 x_{i4} + \varepsilon_i,\\
    y_i^{(4)} &= 2 + \cos (\pi(x_{i1} + x_{i2})) \sin( \pi(x_{i3} + x_{i4}) ) + \varepsilon_i,\\
    y_i^{(5)} &= 1 + \varepsilon_i,
\end{align}
where $i = 1, \ldots, N$, and where $\varepsilon_i$ are $N$ independent draws of a normal random variable with mean 0 and standard deviation $0.1$. In the first four populations, the variable of interest is linked to the auxiliary variables. In the first two populations the link is correctly specified by an AM, even a linear model in population 1. In populations 3 and 4 the AM is not a valid representation of the truth, while in the last population there is no link between the variable of interest and the auxiliary variables.

Two different sampling designs were used for the selection of samples: simple random sampling without replacement (SRSWOR) and stratified sampling (SS). For simple random sampling, a sampling rate of $f = 0.2$ was considered. For stratified sampling, strata were created as follows. First, units were classified into two groups, depending whether their value $x_{i1}$ is larger than the median of $x_1$ or not. In each group created, units were then subdivided into two other groups, depending on whether their value $x_{i2}$ is larger than the median of $x_2$ in each group or not. The procedure was repeated for variables $x_3$ and $x_4$. This resulted in creating 16 strata of size 625 that are somewhat homogeneous with respect to the auxiliary variables. Then, SRSWOR was applied within strata with a sampling rate of $f = 0.2$ in each stratum.

The response probabilities were obtained from
\begin{align}
    p_i = \frac{\exp\left( b_0 + b_1 x_{i1} \right)}{ 1 + \exp\left( b_0 + b_1 x_{i1} \right)},
\end{align}
where $b_0$ and $b_1$ were set to obtain an overall mean response rate which is approximately $75\%$.

One thousand simulations were then conducted as follow. For each simulation, a sample $S$ was selected according to either SRSWOR or SS. For each sample $S$ selected, a respondents set $S_r$ and a nonrespondents set $S_m$ were then created by generating a response indicator vector $\left( r_i | i \in S\right)^\top$, where $r_i$, $i \in S$, was generated from a Bernoulli distribution with parameter $p_i$. Then, for each set of respondents and of nonrespondents obtained, the missing $y_i$, $i \in S_m$, were replaced with imputed $y_i^*$ using the five following imputation methods:
\begin{itemize}
    \item   \textbf{Regression imputation:}  Imputed values $y_i^*$, $i \in S_m$, are obtained by
            \begin{align}
                    y_i^* = \widehat{\beta}_0 + \sum_{j = 1}^q \widehat{\beta}_j x_{ij},
            \end{align}
            where $\widehat{\betab} = (\widehat{\beta}_0,\widehat{\beta}_1,\ldots,\widehat{\beta}_q)^\top$ is defined by
            \begin{align}
                \widehat{\betab} = \left( \sum_{j \in S_r} d_j
                        (1,\xb_j)^\top
                        (1,\xb_j) \right)^{-1} \sum_{i \in S_r} d_i (1,\xb_i)^\top y_i.
            \end{align}
            Regression imputation is based on imputation model model~\ref{imp:mod:0} with $f(x_{i1}, x_{i2},\ldots,x_{iq}) = \beta_0 + \sum_{j = 1}^q \beta_j x_{ij}$.
    \item   \textbf{Mean imputation:} The missing $y_i$, $i \in S_m$, are replaced by the respondents' mean value, that is the imputed values $y_i^*$, $i \in S_m$, are obtained by
            \begin{align}
                y_i^* = \frac{1}{ \sum_{j \in S_r} d_j} \sum_{k \in S_r} d_k y_k.
            \end{align}
           Mean imputation is a particular case of regression imputation where only a constant covariate is considered. It is based on imputation model~\ref{imp:mod:0} with $f(x_{i1}, x_{i2},\ldots,x_{iq}) = \beta_0$.
    \item   \textbf{Nearest neighbor imputation:} The missing $y_i$, $i \in S_m$, are replaced by their respective nearest neighbor in the complete data. The proximity is quantified through the auxiliary variables. Imputed values $y_i^*$, $i \in S_m$, are obtained by
            \begin{align}
                y_i^* = y_{j(i)} \quad \mbox{where} \quad d(\xb_i,\xb_{j(i)}) = \min_{j \in S|r_j = 1} d(\xb_i, \xb_j) ,
            \end{align}
            where $d(\cdot,\cdot)$ is the Euclidean distance.
    \item   \textbf{Random forest imputation:} The missing values were imputed with the nonparametric imputation method using random forest of~\cite{ste:bue:12}. Imputation was carried out using function \texttt{missForest} of \texttt{R} package \texttt{missForest}~\citep{ste:13}. Function \texttt{missForest} begins with an initial guess for the missing values. Then, it sorts the variables according to the amount of missing values starting with the lowest amount. In our case, variable $y$ is last since it is the only one with missing values. The missing values are imputed by first fitting a random forest to the observed values $(y_i, \xb_i)$, $i \in S_r$; then imputing the missing values $y_i$, $i \in S_m$ by applying the trained random forest to $\xb_i$, $i \in S_m$. The procedure is repeated until a stopping criterion is met.
    \item   \textbf{AM imputation:} An AM was fitted using the complete data $(y_i, \xb_i)$, $i \in S_r$, and imputed values $y_i^*$, $i \in S_m$, were obtained through predictions with this model, as explained in Section~\ref{section:method}. Survey weights were considered in the smoothing spline estimator computation of each term, as in the PLS equation of expression~\eqref{equation:pls:pi}. The  model was fitted using function \texttt{gam} of \texttt{R} package \texttt{mgcv}~\citep{woo:14}. Function \texttt{gam} uses $m = 2$ and thin plate splines basis by default. The model is fitted by penalized likelihood maximization and the smoothing parameter is selected by generalized cross validation.
\end{itemize}
The imputed total estimator $\widehat{Y}_I$ was computed for each method and each simulation. Note that all the considered imputation methods use auxiliary information when computing imputed values, except mean imputation.

Moreover, one thousand simulations were conducted to test the accuracy of the bootstrap procedures presented in Section~\ref{section:variance} to estimate the variance of the total. SRSWOR and SS were considered. For each simulation, a sample $S$, a set of respondents $S_r$ and of nonrespondents $S_m$ were created as described above. The missing values were replaced with imputed values using AM imputation. The imputed total estimator $\widehat{Y}_I$ and its bootstrap variance $\V_{boot}(\widehat{Y}_I)$ were computed for each simulation. For the bootstrap variance under SRSWOR, procedure~\ref{procedure:1} was applied where, in step 1, the sample was replicated $k= 1/f = 5$ times to create a pseudopopulation of size 10000 and $B = 100$ bootstrap replicates were generated. For the bootstrap variance under SS, procedure~\ref{procedure:2} was applied where, in step 1, a sample of size 125 was selected in each stratum, that is $n_h' = f \cdot n_h = 125$ for each stratum $h$. This results in integer $n_h'$ and $k_h$ for each stratum $h$.

\subsection{Setting 2: real data}\label{section:setting:2}

We consider the data from the 1992 family expenditure survey (FES), see~\cite{cso:93}. The data is made available by the UK data archive at the University of Essex. To test our method, we considered that the households having a non-missing and larger than zero disposable income (disposable income and self-supply and in kind) of the 1992 FES form the population of interest. The size of this population is $N = 7409$. The variable disposable income was modified as follows. First, it was divided by its mean value. Because income distributions are often right skewed, the natural logarithm of the obtained value plus one was computed. One was added before computing the logarithm to avoid negative values. We suppose that the aim of the survey is to estimate the population total of the modified disposable income. The population was stratified into 12 regions and simple random sampling with a sampling rate of $f = 0.2$ was applied within each region (stratum). The sample size was randomly rounded for 8 strata for which this sampling rate led to a non-integer sample size. For each sampled household, we supposed that the following characteristics were observed:
\begin{itemize}
    \itemsep0em
    \item[$x_{i1}$:] number of adults in household $i$,
    \item[$x_{i2}$:] number of children in household $i$,
    \item[$x_{i3}$:] number of persons economically active in household $i$,
    \item[$x_{i4}$:] age of the head of household $i$,
    \item[$x_{i5}$:] age of the chief economic supporter of household $i$.
\end{itemize}

Such variables could for instance come from a register. It was supposed that the willingness of a household to respond depends on the number of adults in this household and that the households respond independently from each other. Hence, the response probabilities were obtained from
\begin{align}
    p_i = \frac{\exp\left( b_0 + b_1 x_{i1} \right)}{ 1 + \exp\left( b_0 + b_1 x_{i1} \right)},
\end{align}
where $b_0$ and $b_1$ were set to obtain an overall mean response rate which is approximately $70\%$. Then, for each sampled household, a response indicator was generated from a Bernoulli distribution with parameter $p_i$. The modified disposable income was then recorded for respondents and erased for nonrespondents. One thousand simulations were conducted. The same imputation methods as in Section~\ref{section:setting:1} were considered.

Moreover, one thousand simulations were conducted to test the accuracy of the bootstrap procedures presented in section~\ref{section:variance} to estimate the variance of the total. For each simulation, a sample and a set of respondents and of nonrespondents were created as described above. The missing values were replaced with imputed values using AM imputation. The imputed total estimator $\widehat{Y}_I$ and its bootstrap variance $\V_{boot}(\widehat{Y}_I)$ were computed for each simulation. For the bootstrap variance, procedure~\ref{procedure:2} was applied with $B = 100$ bootstrap replicates. We set $n_h' = f \cdot n_h$ and a randomization was applied to round the non-integer $n_h'$ and the non-integer $k_h$~\citep[see][]{sit:92b}.

\subsection{Measures of comparison}\label{section:comparison}

For each simulation and each imputation method of both settings, the population total for the variable of interest was estimated through the imputed estimator of expression~\eqref{equation:imputed:estimator}. To compare the performance of the methods, four comparison measures were recorded. First, to quantify the accuracy of imputed values, the Monte Carlo mean relative prediction error was computed, which is defined as
\begin{align}
    \mbox{MRPE} = \frac{1}{L} \sum_{\ell = 1}^L  \frac{1}{n_m^{(\ell)}} \sum_{i \in S_m^{(\ell)}} \left| \frac{ {y_i^*}^{(\ell)} - y_i}{y_i} \right|,
\end{align}
where $S_m^{(\ell)}$ is the nonrespondents set obtained at the $\ell$-th simulation, $n_m^{(\ell)}$ is the size of $S_m^{(\ell)}$, ${y_i^*}^{(\ell)}$ is the imputed value obtained for $i \in S_m^{(\ell)}$ at the $\ell$-th simulation, and $L$ represents the number of simulations. Then, for each imputation method, the performance of the imputed estimator of expression~\eqref{equation:imputed:estimator} was studied through three comparison measures, namely
\begin{itemize}
    \item   the Monte Carlo relative bias (RB) defined as
            \begin{align}
                \mbox{RB} = \frac{B}{Y},
            \end{align}
            where $\mbox{B} = \widehat{Y}_I^{(\cdot)} - Y$, $\widehat{Y}_I^{(\cdot)}$ represents the mean imputed estimator over the $L$ simulations
            \begin{align}
                \widehat{Y}_I^{(\cdot)} = \frac{1}{L}\sum_{\ell = 1}^L \widehat{Y}_I^{(\ell)},
            \end{align}
            and $\widehat{Y}_I^{(\ell)}$ is the imputed estimator $\widehat{Y}_I$ obtained at the $\ell$-th simulation,
    \item   the Monte Carlo relative root variance (or relative standard deviation) defined as
            \begin{align}
                \mbox{RRVAR} = \frac{\left( \mbox{VAR}\right)^{1/2}}{Y},
            \end{align}
            where
            \begin{align}
                \mbox{VAR} =  \frac{1}{L - 1} \sum_{\ell = 1}^L \left( \widehat{Y}_I^{(\ell)} - \widehat{Y}_I^{(\cdot)} \right)^2 ,
            \end{align}
    \item   the Monte Carlo relative root mean square error defined as
            \begin{align}
                \mbox{RRMSE} = \frac{\left(\mbox{B}^2 + \mbox{VAR} \right)^{1/2}}{Y}.
            \end{align}
\end{itemize}
For AM imputation, the following measures were computed to test the accuracy of the bootstrap variance estimator:
\begin{itemize}
    \item   The Monte Carlo variance of the total estimator:
            \begin{align}
                \mbox{VAR} =  \frac{1}{L - 1} \sum_{\ell = 1}^L \left( \widehat{Y}_I^{(\ell)} - \widehat{Y}_I^{(\cdot)} \right)^2 ,
            \end{align}
    \item   The Monte Carlo expectation of the bootstrap variance estimator:
            \begin{align}
                \mbox{VAR}_{boot} = \frac{1}{L} \sum_{\ell = 1}^L \V_{boot}^{(\ell)}(\widehat{Y}_I),
            \end{align}
            where $\V_{boot}^{(\ell)}(\widehat{Y}_I)$ is the bootstrap variance $\V_{boot}(\widehat{Y}_I)$ obtained at the $\ell$-th simulation,
    \item   The coverage rate CR: the proportion of times the true total $Y$ falls into the 95\% confidence interval
            \begin{align}
                \widehat{Y}_I \pm 1.96 \sqrt{\V_{boot}(\widehat{Y}_I)}.
            \end{align}
\end{itemize}

\subsection{Results of setting 1}\label{section:results:setting:1}

\begin{figure}
\begin{center}
\includegraphics[width=1\textwidth]{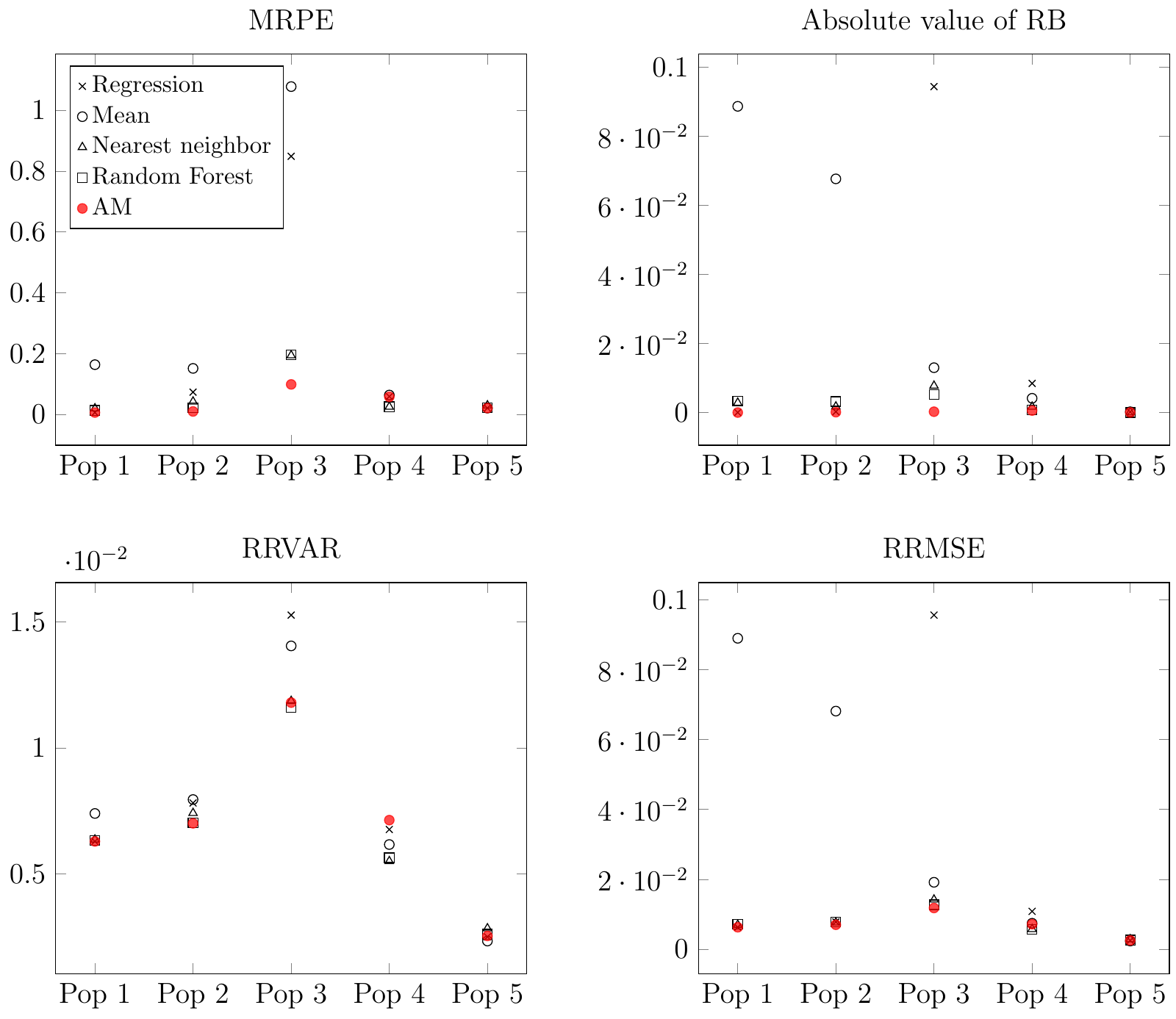}
\caption{Comparison measures of five imputation methods in five populations under SRSWOR.}
\label{fig:srs}
\end{center}
\end{figure}

\begin{figure}
\begin{center}
\includegraphics[width=1\textwidth]{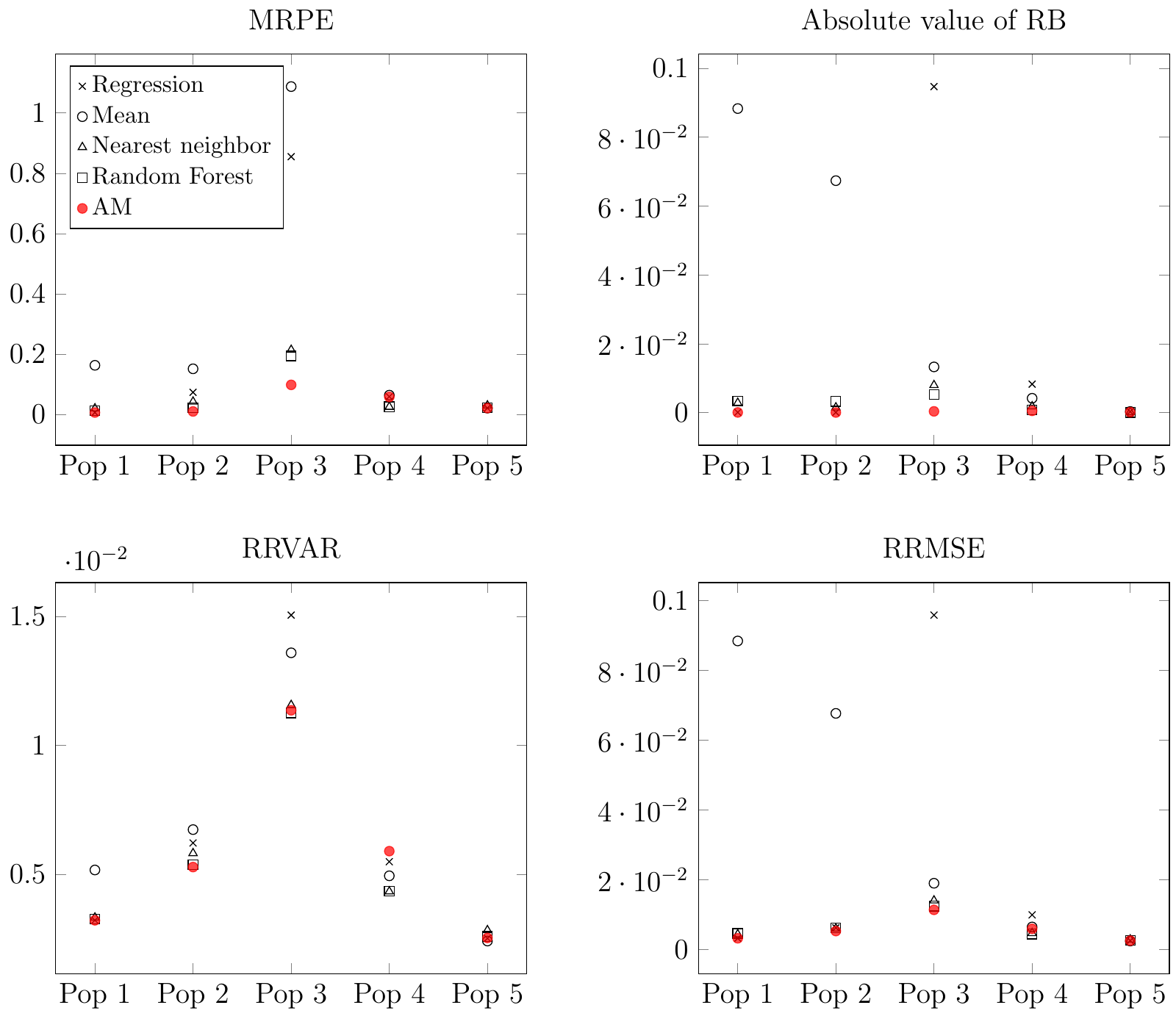}
\caption{Comparison measures of five imputation methods in five populations under SS.}
\label{fig:ss}
\end{center}
\end{figure}

Figure~\ref{fig:srs}, Figure~\ref{fig:ss}, and Table~\ref{table:variance} display the results of Setting 1. Table~\ref{table:ranks} reports the average ranks over the populations of each imputation method for each measure of comparison. The absolute value of RB was considered.

\begin{table}[htb]\caption{Average ranks over five populations of each imputation method for each measure of comparison (in absolute value). \label{table:ranks}}
\begin{center}
\begin{tabular}{lcccc}
\toprule
Imputation method           & \multicolumn{1}{c}{MRPE} & \multicolumn{1}{c}{RB} & \multicolumn{1}{c}{RRVAR} &\multicolumn{1}{c}{RRMSE}  \\
\midrule
&\multicolumn{4}{c}{Simple random sampling (SRSWOR)}\\
            \cmidrule{2-5}
Regression                  &   3.0&    3.6&    3.4&    3.6\\
Mean                        &   4.2&    4.6&    3.6&    3.8\\
Nearest Neighbor            &   3.0&    3.0&    3.2&    3.0\\
Random Forest               &   2.8&    2.6&    2.4&    2.8\\
AM                          &   2.0&    1.2&    2.4&    1.8\\
\cmidrule{2-5}
&\multicolumn{4}{c}{Stratified sampling (SS)}\\
\cmidrule{2-5}
Regression                  &   3.0&    3.6&    3.4&    3.6\\
Mean                        &   4.2&    4.6&    3.6&    3.8\\
Nearest Neighbor            &   3.2&    3.0&    3.2&    3.0\\
Random Forest               &   2.6&    2.6&    2.4&    2.8\\
AM                          &   2.0&    1.2&    2.4&    1.8\\
\bottomrule
\end{tabular}
\end{center}
\end{table}

\begin{table}[htb]\caption{Monte Carlo variance of the total, Monte carlo expectation of the bootstrap variance and coverage rate associated with AM imputation for two different sampling designs and five populations.\label{table:variance}}
\begin{center}
\begin{tabular}{lccc}
\toprule
& \multicolumn{1}{l}{VAR} & \multicolumn{1}{l}{$\mbox{VAR}_{boot}$} & \multicolumn{1}{c}{CR}  \\
\midrule
&\multicolumn{3}{c}{Simple random sampling (SRSWOR)}\\
\cmidrule{2-4}
Population 1            &   91033.21&    90995.88&    0.95\\
Population 2            &   39388.03&    40340.57&    0.95\\
Population 3            &   24409.33&    23382.72&    0.94\\
Population 4            &   15566.86&    14537.13&    0.94\\
Population 5            &   597.74  &      605.67&    0.96\\
\cmidrule{2-4}
&\multicolumn{3}{c}{Stratified sampling (SS)}\\
\cmidrule{2-4}
Population 1            &   25176.60&    23171.71&    0.94\\
Population 2            &   23966.30&    24363.67&    0.95\\
Population 3            &   22227.70&    21810.21&    0.95\\
Population 4            &   11461.04&    10965.36&    0.93\\
Population 5            &     643.41&      600.36&    0.93\\
\bottomrule
\end{tabular}
\end{center}
\end{table}

We first comment the results shown in Figures~\ref{fig:srs} and \ref{fig:ss}. When  functional dependence between the variable of interest and the auxiliary variables is additive (populations 1 and 2), AM imputation provides the best results. If, moreover, this functional dependence is linear (population 1), regression imputation performs as well as AM imputation. When there is no dependence between the variable of interest and the auxiliary variables (population 5), all five methods perform fairly similarly. 
Because the functional dependence between the variable of interest and the auxiliary variables is not additive in populations 3 and 4, the results for these two populations allow us to study the performance of AM imputation under model misspecification. We can see that AM imputation still performs the best overall in population 3, except for the RRVAR, which is slightly smaller for random forest. The reason for the good performance of AM imputation in this population is that, even though the functional dependence is not additive, it can be well approximated by an additive function. In population 4, the situation is less obvious and it is difficult to rank the imputation methods. It seems that, in this population, nearest neighbor and random forest perform slightly better than the other methods. In order to produce a global index of performance we ranked the imputing methods for each population and each performance criterion. The results, reported in Table~\ref{table:ranks} show that, globally, AM imputation performs better than the other imputation methods considered.

The performance of the bootstrap-based estimators of the variance is assessed in Table~\ref{table:variance}. Whether the functional dependence between the variable of interest and the auxiliary variables is additive (populations 1 and 2) or not (populations 3, 4, 5), the bootstrap variance is very close to the variance obtained by simulation. Also, it leads to very good coverage rates (between 93\% and 96\%) across all five populations considered.

\subsection{Results of setting 2}\label{section:results:setting:2}

Table~\ref{table:fes:total} and Table~\ref{table:fes:variance} display the results of our analysis performed under setting 2. The numbers in brackets in Table~\ref{table:fes:total} report the ranks of each imputation method for each measure of comparison.

\begin{table}[htb]\caption{Comparison measures for five imputation methods for FES data. \label{table:fes:total}}
\begin{center}
\small
\begin{tabular}{lcccc}
\toprule
Imputation method  & \multicolumn{1}{c}{MRPE} & \multicolumn{1}{c}{RB} & \multicolumn{1}{c}{RRVAR} &\multicolumn{1}{c}{RRMSE}  \\
                   & \multicolumn{1}{c}{$\times 10^{1}$} & \multicolumn{1}{c}{$\times 10^{-2}$}  & \multicolumn{1}{c}{$\times 10^{-2}$} &\multicolumn{1}{c}{$\times 10^{-2}$}  \\
\midrule
Regression                 &  3.37(3)&    0.76(3)&   1.45(3)&   1.64(3)\\
Mean                       &  4.63(5)&    5.51(5)&   1.56(5)&   5.73(5)\\
Nearest Neighbor           &  3.45(4)&    0.81(4)&   1.54(4)&   1.74(4)\\
Random Forest              &  3.06(2)&    0.19(2)&   1.41(1)&   1.42(2)\\
AM                         &  2.99(1)&    0.05(1)&   1.41(1)&   1.41(1)\\
\bottomrule
\end{tabular}
\end{center}
\end{table}

\begin{table}[htb]\caption{Monte Carlo variance of the total, Monte carlo expectation of the bootstrap variance and coverage rate associated with AM imputation for FES data.\label{table:fes:variance}}
\begin{center}
\begin{tabular}{ccc}
\toprule
 \multicolumn{1}{c}{VAR} & \multicolumn{1}{c}{$\mbox{VAR}_{boot}$} & \multicolumn{1}{c}{CR}  \\
\midrule
   4194.50&    4042.78&    0.94\\
\bottomrule
\end{tabular}
\end{center}
\end{table}

We can see that AM imputation outperforms the competing imputation methods in terms of MRPE and in terms of RB. AM imputation and random forest perform equally and slightly better than the other three methods in terms of RRVAR. With this data, the bootstrap variance yields a coverage rate of 94\% that is close to the theoretically stated value of 95\%.

As we can see from the results of both settings, AM imputation performs the best overall, closely followed by random forest. This is not surprising since random forest is also nonparametric. Two advantage of random forest over our imputation method are: 1) it can handle mixed-type data and 2) auxiliary variables can have missing values. Two advantages of our method are: 1) it is fast and 2) it allows us to take design weights into account in the imputation model.

\section{Conclusion}\label{section:conclusion}

A new imputation method for nonresponse in surveys based on spline smoothing within the additive model paradigm was proposed. The simulations indicate that the new method is very flexible and can capture a large spectrum of functional dependencies between the variable of interest and the auxiliary variables. Since the model requires only weak assumptions, it is less susceptible to model misspecification than other models such as parametric ones. Most importantly, the AM formulation makes it possible to consider several auxiliary variables in the imputation process without running into the curse of dimensionality phenomenon. A bootstrap procedure to estimate the variance of the total under SRSWOR and SS was suggested.

Through a simulation study, the new imputation method was confirmed to perform well in many different situations. The main conclusions of the simulation study are the following. AM imputation performs better than the other imputation methods considered when the functional dependence between the variable of interest and the auxiliary variables is additive or when this dependence can be well approximated by an additive function. When this dependence is not well approximated by an additive function or when there is no dependence between the variable of interest and the auxiliary variables, AM imputation shows a performance similar to that of the other imputation methods considered. In all the cases studied, the proposed bootstrap-based variance estimates were close to the true Monte Carlo variance and produced very good coverage rates.

Future work include extending the current method to situations in which the samples are dependent and improving the computational speed of the variance via parallel processing.

\section*{Acknowledgements}
The authors thank Yves Till\'e for his constructive suggestions. This research was supported by the Swiss National Science Foundation, project number P1NEP2\_151904 (CH) and the Natural Science and Engineering Research Council of Canada (RVC).
\clearpage


\end{document}